\documentclass[zpreprint,zbstpl]{./LaTeX/zeus/zeus_paper}
\usepackage[english]{babel}

\newcommand{\ZcoosysB}{%
The ZEUS coordinate system is a right-handed Cartesian system, with the $Z$
axis pointing in the proton beam direction, referred to as the ``forward
direction'', and the $X$ axis pointing left towards the centre of HERA.
The polar angle, $\theta$, is measured with respect to the proton beam
direction. The coordinate origin is at the nominal interaction point.\xspace}
\newcommand{\Zpsrap}{%
The pseudorapidity is defined as $\eta=-\ln\left(\tan\frac{\theta}{2}\right)$,
where the polar angle, $\theta$, is measured with respect to the proton beam
direction.\xspace}

\newcommand{\ZcoosysfnB}{\footnote{\ZcoosysB}}

\newcommand{\ZcoosysfnBeta}{\footnote{\ZcoosysB\Zpsrap}}

\newcommand{\Zctddesc}[1]{%
Charged particles are tracked in the central tracking detector (CTD)~\citeCTD,
which operates in a magnetic field of $1.43\Tesla$ provided by a thin 
superconducting solenoid. The CTD consists of 72~cylindrical drift chamber 
layers, organised in nine~superlayers covering the polar-angle\ZcoosysfnBeta~region 
\mbox{$15^\circ<\theta<164^\circ$}. The relative transverse-momentum 
resolution for
full-length tracks is $\sigma(p_T)/p_T=0.0058p_T\oplus0.0065\oplus0.0014/p_T$,
with $p_T$ in $\Gev$.\xspace\\
In 2001 a silicon microvertex detector (MVD)~\cite{nim:a453:89,nim:a505:663} was installed 
inside the CTD. The MVD is organised into a barrel with 3 cylindrical layers and a forward 
section with four planar layers perpendicular to the HERA beam direction. The barrel 
contains 600 single-sided silicon strip sensors each having 512 120 $\rm{\mu m}$ wide strips; the 
forward section contains 112 sensors each of which has 480 120 $\rm{\mu m}$ wide strips.\xspace\\}

\chardef\usc=95
\chardef\til=126
\catcode`\@=11 
\DeclareRobustCommand\xdotspace{\futurelet\@let@token\@xdotspace}
\def\@xdotspace{%
  \ifx\@let@token.\else
  \ifx\@let@token\bgroup.\else
  \ifx\@let@token\egroup.\else
  \ifx\@let@token\/.\else
  \ifx\@let@token\ .\else
  \ifx\@let@token~.\else
  \ifx\@let@token!.\else
  \ifx\@let@token,.\else
  \ifx\@let@token:.\else
  \ifx\@let@token;.\else
  \ifx\@let@token?.\else
  \ifx\@let@token/.\else
  \ifx\@let@token'.\else
  \ifx\@let@token).\else
  \ifx\@let@token-.\else
  \ifx\@let@token\@xobeysp.\else
  \ifx\@let@token\space.\else
  \ifx\@let@token\@sptoken.\else
   .\space
   \fi\fi\fi\fi\fi\fi\fi\fi\fi\fi\fi\fi\fi\fi\fi\fi\fi\fi}
\catcode`\@=12 

\newcommand{\stru}[2]{%
   \relax\ifmmode\hbox{\vrule height#1 depth#2 width0pt}%
   \else\vrule height#1 depth#2 width0pt\fi}

\newcommand{\Ronum}[1]{\uppercase\expandafter{\romannumeral#1}}
\newcommand{\ronum}[1]{\expandafter{\romannumeral#1}}
\DeclareRobustCommand{\LaTeXZ}{%
  \LaTeX\kern-.05em4\kern-.1em
  {\raisebox{-0.2ex}{$\scriptstyle\text{ZEUS}$}}\xspace}

\DeclareMathAlphabet{\mathbf}{OT1}{cmr}{bx}{sl}
\newcommand{\eVdist}{\kern-0.06667em}

\newcommand{\Gev}{{\text{Ge}\eVdist\text{V\/}}}

\newcommand{\gev}{{\,\text{Ge}\eVdist\text{V\/}}}

\newcommand{\pb}{\,\text{pb}}

\newcommand{\Tesla}{\,\text{T}}

\newcommand{\slashfrac}[2]{%
  \raisebox{0.5ex}{\ensuremath #1}\kern-0.12em/\kern-0.08em
  \raisebox{-.8ex}{\ensuremath #2}}

\newcommand{\sqr}[3]{%
    {\vcenter{\hrule height.#3ex\hbox{\vrule width.#2ex height#1ex
     \kern#1ex\vrule width.#3ex}\hrule height.#2ex}}}

\catcode`\@=11 
\newcommand{\parenbar}{\mathpalette\p@renb@r}
\def\p@renb@r#1#2{\vbox{%
  \ifx#1\scriptscriptstyle \dimen@.7em\dimen@ii.2em\else
  \ifx#1\scriptstyle \dimen@.8em\dimen@ii.25em\else
  \dimen@1em\dimen@ii.4em\fi\fi \offinterlineskip
  \ialign{\hfill##\hfill\cr
    \vbox{\hrule width\dimen@ii}\cr
    \noalign{\vskip-.3ex}%
    \hbox to\dimen@{$\mathchar300\hfil\mathchar301$}\cr
    \noalign{\vskip-.3ex}%
    $#1#2$\cr}}}
\catcode`\@=12 

\newcommand{\IP}{{\rm I$\kern-0.01667em$P}\xspace}

\mathchardef\qsm=63
\mathchardef\pls=43
\mathchardef\mns=512
\mathchardef\plm=518
\mathchardef\eql=61
\mathchardef\smallleft=300
\mathchardef\smallright=301
\mathchardef\les=316
\mathchardef\gre=318
\mathchardef\leq=532
\mathchardef\grq=533

\catcode`\@=11 
\newcounter{pict@width}
\newcounter{pict@height}
\newlength{\pict@scale}
\newcommand{\psfigadd}[4]{%
\setcounter{pict@width}{1*\ratio{#2+\pict@scale/2}{\pict@scale}}
\setcounter{pict@height}{1*\ratio{#3+\pict@scale/2}{\pict@scale}}
\setlength{\unitlength}{\pict@scale}
\hbox to #2{\hspace{-\fill}\begin{picture}(\thepict@width,\thepict@height)
\put(0,0){\psfig{figure=#1,width=#2,height=#3,clip=}}
\SetScale{0.283466457}
\SetWidth{1.763889}
{#4}
\end{picture}}
}
\newcounter{pict@widthfst}
\newcounter{pict@widthscd}
\newcounter{pict@widthtot}
\newcommand{\psfigaddtwo}[7]{%
\setcounter{pict@widthfst}{1*\ratio{#2+\pict@scale/2}{\pict@scale}}
\setcounter{pict@widthscd}{1*\ratio{#2+#4+\pict@scale/2}{\pict@scale}}
\setcounter{pict@widthtot}{1*\ratio{#2+#4+#6+\pict@scale/2}{\pict@scale}}
\setcounter{pict@height}{1*\ratio{#3+\pict@scale/2}{\pict@scale}}
\setlength{\unitlength}{\pict@scale}
\hbox{\hspace{-\fill}\begin{picture}(\thepict@widthtot,\thepict@height)
\put(0,0){\psfig{figure=#1,width=#2,height=#3,clip=}}
\put(\thepict@widthscd,0){\psfig{figure=#5,width=#6,height=#3,clip=}}
\SetScale{0.283466457}
\SetWidth{1.763889}
{#7}
\end{picture}}
}
\newcommand{\psfigror}[4]{%
\setcounter{pict@width}{1*\ratio{#2+\pict@scale/2}{\pict@scale}}
\setcounter{pict@height}{1*\ratio{#3+\pict@scale/2}{\pict@scale}}
\setlength{\unitlength}{\pict@scale}
\hbox{\begin{picture}(\thepict@width,\thepict@height)
\put(0,\thepict@height){\psfig{figure=#1,width=#3,height=#2,clip=,angle=270}}
\SetScale{0.283466457}
\SetWidth{1.763889}
{#4}
\end{picture}}
}
\newcommand{\psfigrol}[4]{%
\setcounter{pict@width}{1*\ratio{#2+\pict@scale/2}{\pict@scale}}
\setcounter{pict@height}{1*\ratio{#3+\pict@scale/2}{\pict@scale}}
\setlength{\unitlength}{\pict@scale}
\hbox{\begin{picture}(\thepict@width,\thepict@height)
\put(0,0){\psfig{figure=#1,width=#3,height=#2,clip=,angle=90}}
\SetScale{0.283466457}
\SetWidth{1.763889}
{#4}
\end{picture}}
}
\catcode`\@=12 
\newlength\listtextwidth

\catcode`\@=11 
\newlength{\@tabfninsert}
\newlength{\@tabfnwidth}
\newcommand{\tabfootnote}[2]{%
  \setlength{\@tabfninsert}{0.8em}
  \setlength{\@tabfnwidth}{\textwidth}
  \addtolength{\@tabfnwidth}{-\@tabfninsert}
  \addtolength{\@tabfnwidth}{-0.4em}
  \noindent\makebox[\@tabfninsert][r]{\footnotesize$^{#1}$\hfil}\hfill%
  \parbox[t]{\@tabfnwidth}{\footnotesize #2\hfill}}
\catcode`\@=12 

\newcommand{\PTM}       {P_{T,{\rm miss}}}

\def\citeCTD{{\cite{%
nim:a279:290,*npps:b32:181,*nim:a338:254%
}}\xspace}
\def\citeCAL{{\cite{%
nim:a309:77,*nim:a309:101,*nim:a321:356,*nim:a336:23%
}}\xspace}

\begin{document}
\title{
Measurement of high-\boldmath{$Q^2$} deep inelastic scattering cross sections 
with a longitudinally polarised positron beam at HERA
}                                                       
                    
\author{ZEUS Collaboration}
\prepnum{DESY-06-015}

\abstract{
The cross sections for charged and neutral current 
deep inelastic scattering in $e^{+}p$ collisions with a longitudinally polarised 
positron beam have been measured using the ZEUS detector at HERA.
The results, based on data corresponding to an integrated luminosity of $23.8 \pb^{-1}$ at
 $\sqrt{s}=318 \gev$, are given for both $e^+p$ charged current and neutral current deep 
inelastic scattering for both positive and negative values of the longitudinal 
polarisation of the positron beam.
Single differential cross sections are presented for the kinematic region $Q^{2}>200 \gev^{2}$. 
The measured cross sections are compared to 
the predictions of the Standard Model. A fit to the data yields
\mbox{$\sigma^{\rm CC}(P_{e}=-1)=7.4 \pm 3.9 ({\rm stat.}) \pm 1.2 ({\rm syst.})~{\rm pb}$},
 which is consistent within two standard deviations with the absence of right-handed charged currents in the Standard Model.

\vspace{2cm}
\begin{center}
{\it Dedicated to our friend and colleague Nikolaj Pavel.}
\end{center}

}

\makezeustitle

\def\3{\ss}                                                                                        
\pagenumbering{Roman}                                                                              
                                                   %
\begin{center}                                                                                     
{                      \Large  The ZEUS Collaboration              }                               
\end{center}                                                                                       
  S.~Chekanov,                                                                                     
  M.~Derrick,                                                                                      
  S.~Magill,                                                                                       
  S.~Miglioranzi$^{   1}$,                                                                         
  B.~Musgrave,                                                                                     
  D.~Nicholass$^{   1}$,                                                                           
  \mbox{J.~Repond},                                                                                
  R.~Yoshida\\                                                                                     
 {\it Argonne National Laboratory, Argonne, Illinois 60439-4815}, USA~$^{n}$                       
\par \filbreak                                                                                     
  M.C.K.~Mattingly \\                                                                              
 {\it Andrews University, Berrien Springs, Michigan 49104-0380}, USA                               
\par \filbreak                                                                                     
  N.~Pavel~$^{\dagger}$, A.G.~Yag\"ues Molina \\                                                   
  {\it Institut f\"ur Physik der Humboldt-Universit\"at zu Berlin,                                 
           Berlin, Germany}
% ...
\newpage                                                                                           
$^{\    1}$ also affiliated with University College London, UK \\                                  
% ...
$^{\    6}$ supported by the research grant no. 1 P03B 04529 (2005-2008) \\                        
$^{\    7}$ now at DESY group FEB, Hamburg, Germany \\                                             
$^{\    8}$ now at Argonne National Laboratory, Argonne, IL, USA \\                                
$^{\    9}$ also at INP, Cracow, Poland \\                                                         
$^{  10}$ on leave of absence from FPACS, AGH-UST, Cracow, Poland \\                               
$^{  11}$ partly supported by Moscow State University, Russia \\                                   
$^{  12}$ also affiliated with DESY \\                                                             
$^{  13}$ now at ICEPP, University of Tokyo, Japan \\                                              
$^{  14}$ also at University of Tokyo, Japan \\                                                    
$^{  15}$ Ram{\'o}n y Cajal Fellow \\                                                              
% ...
$^{\dagger}$ deceased \\                                                                           
%
\newpage   
                                                           %
                                                           %
\begin{tabular}[h]{rp{14cm}}                                                                       
$^{a}$ &  supported by the Natural Sciences and Engineering Research Council of Canada (NSERC) \\  
$^{b}$ &  supported by the German Federal Ministry for Education and Research (BMBF), under        
          contract numbers HZ1GUA 2, HZ1GUB 0, HZ1PDA 5, HZ1VFA 5\\                                
$^{c}$ &  supported in part by the MINERVA Gesellschaft f\"ur Forschung GmbH, the Israel Science   
          Foundation (grant no. 293/02-11.2) and the U.S.-Israel Binational Science Foundation \\  
$^{d}$ &  supported by the German-Israeli Foundation and the Israel Science Foundation\\           
$^{e}$ &  supported by the Italian National Institute for Nuclear Physics (INFN) \\                
$^{f}$ &  supported by the Japanese Ministry of Education, Culture, Sports, Science and Technology 
          (MEXT) and its grants for Scientific Research\\                                          
$^{g}$ &  supported by the Korean Ministry of Education and Korea Science and Engineering          
          Foundation\\                                                                             
$^{h}$ &  supported by the Netherlands Foundation for Research on Matter (FOM)\\                   
$^{i}$ &  supported by the Polish State Committee for Scientific Research, grant no.               
          620/E-77/SPB/DESY/P-03/DZ 117/2003-2005 and grant no. 1P03B07427/2004-2006\\             
$^{j}$ &  partially supported by the German Federal Ministry for Education and Research (BMBF)\\   
$^{k}$ &  supported by RF Presidential grant N 1685.2003.2 for the leading scientific schools and  
          by the Russian Ministry of Education and Science through its grant for Scientific        
          Research on High Energy Physics\\                                                        
$^{l}$ &  supported by the Spanish Ministry of Education and Science through funds provided by     
          CICYT\\                                                                                  
$^{m}$ &  supported by the Particle Physics and Astronomy Research Council, UK\\                   
$^{n}$ &  supported by the US Department of Energy\\                                               
$^{o}$ &  supported by the US National Science Foundation\\                                        
$^{p}$ &  supported by the Polish Ministry of Scientific Research and Information Technology,      
          grant no. 112/E-356/SPUB/DESY/P-03/DZ 116/2003-2005 and 1 P03B 065 27\\                  
$^{q}$ &  supported by FNRS and its associated funds (IISN and FRIA) and by an Inter-University    
          Attraction Poles Programme subsidised by the Belgian Federal Science Policy Office\\     
$^{r}$ &  supported by the Malaysian Ministry of Science, Technology and                           
Innovation/Akademi Sains Malaysia grant SAGA 66-02-03-0048\\                                       
\end{tabular}                                                                                      
                                                           %
                                                           %

\pagenumbering{arabic} 
\pagestyle{plain}

\section{\bf Introduction}
\label{s:intro}

Deep inelastic scattering (DIS) of leptons off nucleons is an important process in the
understanding of the structure of the proton and has been vital in the development of the 
Standard Model (SM). The HERA $ep$ collider allows
the exploration of DIS at high values of the negative four-momentum-transfer squared, $Q^{2}$. 
Using data taken in the years 1994-2000 the H1 and ZEUS collaborations have reported
measurements of the cross sections for charged current (CC)
and neutral current (NC) DIS~\mcite{pl:b324:241,*zfp:c67:565,*pl:b379:319,*np:b470:3,*np:b497:3,
*epj:c13:609,*epj:c19:269,*epj:c21:33,*epj:c30:1,prl:75:1006,zfp:c72:47,epj:c11:427,epj:c12:411,
epj:c21:443,pl:b539:197,epj:c28:175,epj:c32:1,pr:d70:052001-tmp-43d0e0ea}. 
These measurements extend the 
kinematic region covered by fixed-target experiments~\mcite{zfp:c25:29,*zfp:c49:187,*zfp:c53:51,
*zfp:c62:575} to higher $Q^{2}$ values and probe the electroweak 
sector of the Standard Model. 

Polarised electron-nucleon 
deep inelastic scattering was 
first performed in the 1970s at low values of $Q^{2}$.
The results established parity violation attributable to the weak neutral current~\cite{plb:77:347-tmp-43e37e20}.
Since 2002, the upgraded HERA collider has delivered longitudinally polarised lepton beams to the 
collider experiments. The luminosity was also higher than in previous years.
In the kinematic range of HERA, the SM predicts that the cross sections for charged and
neutral current $ep$ DIS should exhibit specific dependencies on the longitudinal
polarisation of the incoming lepton beam. 
The absence of right-handed charged currents leads to the prediction that the charged current 
cross section will be a linear function of polarisation, vanishing for right-handed (left-handed) 
electron (positron) beams.

This paper presents measurements of the cross sections for $e^+ p$ CC and NC DIS
at high $Q^{2}$ with longitudinally polarised positron beams using the ZEUS detector. 
The measurements are based on $11.5 \pb^{-1}$
of data collected between April and June 2004 at a mean luminosity-weighted polarisation of $-0.41$, 
and $12.3 \pb^{-1}$ collected between June and August 2004 at a polarisation of $+0.32$. 
During this time HERA collided protons of energy
$920 \gev$ with positrons of energy $27.5 \gev$, yielding collisions at a centre-of-mass
energy of $318 \gev$. The measured cross sections are compared to the predictions of the SM.
Similar results have recently been published by the H1 Collaboration~\cite{pl:b634:173}.

%
%
\section{\bf Standard Model predictions}
\label{s:Kincross}

Inclusive deep inelastic lepton-proton scattering can be described in terms of the kinematic 
variables $x$ and $Q^2$.
The variable $Q^2$ is defined by $Q^2 = -q^2 = -(k-k')^2$ where $k$ and $k'$ 
are the four-momenta of the incoming and scattered lepton, respectively. Bjorken $x$ is defined
by $x=Q^2/(2P \cdot q)$, where $P$ is the four-momentum of the incoming proton. The inelasticity
variable $y$ is determined from $Q^2=sxy$, where $s$ is the square 
of the lepton-proton centre-of-mass energy (neglecting the masses of the incoming particles).

The electroweak Born level cross section for the CC reaction

\begin{equation}
e^{+}p\rightarrow \bar{\nu}_{e}X, \nonumber
\end{equation}

with a longitudinally polarised positron beam, can be expressed at leading order in QCD as~\cite{devenish:2003:dis}

\begin{equation}
\frac{d^2 \sigma ^{\rm CC} (e^+ p)}{dxdQ^2}=(1+P_{e})\frac{G_{F}^{2}}{4\pi x}\bigg(\frac{M_{W}^{2}}{M_{W}^{2}+Q^{2}}\bigg) ^{2} \cdot \bigg[ Y_{+}F_{2}^{\rm CC}(x,Q^{2})-Y_{-}xF_{3}^{\rm CC}(x,Q^{2}) \bigg], \nonumber
\end{equation}

where $G_{F}$ is the Fermi constant, $M_{W}$ is the mass of the $W$ boson and $Y_{\pm}=1\pm(1-y)^{2}$. 
The structure-functions $F_{2}^{\rm{CC}}$ and $xF_{3}^{\rm{CC}}$ may be written in terms of sums and differences of 
quark and antiquark parton density functions (PDFs) of the proton as follows:

\begin{equation}
F_{2}^{\rm CC} = x[d(x,Q^{2})+s(x,Q^{2})+\bar{u}(x,Q^{2})+\bar{c}(x,Q^{2})], \nonumber
\end{equation}
\begin{equation}
xF_{3}^{\rm CC} = x[d(x,Q^{2})+s(x,Q^{2})-\bar{u}(x,Q^{2})-\bar{c}(x,Q^{2})], \nonumber
\end{equation}

where, for example, the PDF $d(x,Q^{2})$ gives the number density of down quarks
with momentum-fraction $x$ at a given $Q^2$. The longitudinal polarisation of the positron beam is defined as

\begin{equation}
P_{e}=\frac{N_{R}-N_{L}}{N_{R}+N_{L}}, \nonumber
\end{equation}
  
where $N_{R}$ and $N_{L}$ are the numbers of right\footnote{At HERA beam energies the mass of the incoming leptons may be neglected, and therefore the difference between handedness and helicity may also be neglected.}- and left-handed positrons in the beam, respectively. Similarly the cross section for the NC reaction

\begin{equation}
e^{+}p\rightarrow e^{+}X, \nonumber
\end{equation}

can be expressed as~\cite{devenish:2003:dis}

\begin{equation}
\frac{d^2 \sigma ^{\rm NC} (e^+ p)}{dxdQ^2}=\frac{2\pi \alpha ^2}{xQ^4} [H_{0}^{+}-P_{e}H_{P_{e}}^{+}], \nonumber
\end{equation}

where $\alpha$ is the QED coupling constant and $H_{0}^{+}$ and $H_{P_{e}}^{+}$ contain the unpolarised and polarised structure functions, respectively, such that at leading order in QCD

 \begin{equation}
 H_{0}^{+}=Y_{+}F_{2}^{0}-Y_{-}xF_{3}^{0}, \hspace{1cm} F_{2}^{0}=\sum_q x(+\bar{q})A_{q}^{0}, \hspace{1cm} xF_{3}^{0}=\sum _q x(q-\bar{q})B_{q}^{0}, \nonumber
 \end{equation}

 and

 \begin{equation}
 H_{P_{e}}^{+}=Y_{+}F_{2}^{P_{e}}-Y_{-}xF_{3}^{P_{e}}, \hspace{1cm} F_{2}^{P_{e}}=\sum_q x(q+\bar{q})A_{q}^{P_{e}}, \hspace{1cm} xF_{3}^{P_{e}}=\sum _q x(q-\bar{q})B_{q}^{P_{e}}, \nonumber
 \end{equation}

 where $q(x,Q^{2})$ and $\bar{q}(x,Q^{2})$ are the quark and antiquark PDFs, respectively, and the sums run over the five active quark flavours. The $A$ and $B$ coefficients contain the quark and positron couplings to the photon and $Z$ boson and are given by

 \begin{equation}
 A_{q}^{0}=e_{q}^{2}-2e_{q}v_{q}v_{e}\chi_{Z}+(v_{q}^{2}+a_{q}^{2})(v_{e}^{2}+a_{e}^{2})\chi_{Z}^{2}, \nonumber
 \end{equation}

 \begin{equation}
 B_{q}^{0}=-2e_{q}a_{q}a_{e}\chi_{Z}+4v_{q}a_{q}v_{e}a_{e}\chi_{Z}^{2}, \nonumber
 \end{equation}

 and

 \begin{equation}
 A_{q}^{P_{e}}=2e_{q}v_{q}a_{e}\chi_{Z}-2(v_{q}^{2}+a_{q}^{2})v_{e}a_{e}\chi_{Z}^{2}, \nonumber
 \end{equation}

 \begin{equation}
 B_{q}^{P_{e}}=2e_{q}a_{q}v_{e}\chi_{Z}-2v_{q}a_{q}(v_{e}^{2}+a_{e}^{2})\chi_{Z}^{2}, \nonumber
 \end{equation}

 where $e_{f}$ is the electric charge in units of the positron charge and $a_{f}$ and $v_{f}$ are the axial and vector couplings of the fermion $f$. The couplings are defined by $a_{f}=I_{3}^{f}$ and $v_{f}=I_{3}^{f}-2e_{f}\sin ^{2}\theta_W$ where $I_{3}^{f}$ is the third component of weak isospin  and $\theta_W$ is the Weinberg angle. The quantity $\chi_{Z}$ is proportional to the ratio of the $Z^{0}$ and photon propagators:

 \begin{equation}
 \chi_{Z}=\frac{1}{\sin ^{2} 2\theta_W} \bigg ( \frac{Q^{2}}{M_{Z}^{2}+Q^{2}} \bigg ), \nonumber
 \end{equation}

 where $M_{Z}$ is the mass of the $Z^{0}$ boson.

\section{\bf Experimental apparatus}
\label{s:detector}

A detailed description of the ZEUS detector can be found 
elsewhere~\cite{zeus:1993:bluebook}. A brief outline of the 
components most relevant for this analysis is given
below.

Charged particles are tracked in the central tracking detector (CTD)~\citeCTD,
which operates in a magnetic field of $1.43\Tesla$ provided by a thin 
superconducting solenoid. The CTD consists of 72~cylindrical drift chamber 
layers, organised in nine~superlayers covering the polar-angle\ZcoosysfnB~region 
\mbox{$15^\circ<\theta<164^\circ$}. 
In 2001 a silicon microvertex detector (MVD)~\cite{nim:a453:89,*nim:a505:663} was installed 
between the beampipe and the inner radius of the CTD. 
The MVD is organised into a barrel with 3 cylindrical layers and a forward 
section with four planar layers perpendicular to the HERA beam direction. The barrel 
contains 600 single-sided silicon strip sensors each having 512 strips of width 120~$\rm{\mu m}$; the 
forward section contains 112 sensors each of which has 480 strips of width 120~$\rm{\mu m}$. 
Charged-particle tracks were reconstructed using information from the CTD and MVD. 

The high-resolution uranium--scintillator calorimeter (CAL)~\citeCAL consists 
of three parts: the forward (FCAL), the barrel (BCAL) and the rear (RCAL)
calorimeter, covering 99.7\% of the solid angle around the nominal interaction point. 
Each part is subdivided transversely into towers and
longitudinally into one electromagnetic section (EMC) and either one (in RCAL)
or two (in BCAL and FCAL) hadronic sections (HAC). The smallest subdivision of
the calorimeter is called a cell.  The CAL relative energy resolutions, 
as measured under
test-beam conditions, are $\sigma(E)/E=0.18/\sqrt{E}$ for electrons and
$\sigma(E)/E=0.35/\sqrt{E}$ for hadrons, with $E$ in $\Gev$. The timing resolution of the CAL is better than 1~ns 
for energy deposits exceeding 4.5~\gev. The position of the interaction vertex along the beam direction can be 
reconstructed from the arrival time of energy deposits in FCAL. The resolution is about 10~cm for events with 
FCAL energy above 25~\gev, improving to about 8~cm for FCAL energy above 100~\gev.

An iron structure that surrounds the CAL is instrumented as a backing
calorimeter (BAC)~\cite{nim:a313:126} to measure energy leakage from the CAL. Muon chambers in the forward, 
barrel and rear~\cite{nim:a333:342} regions are used in this analysis to veto background events induced by 
cosmic-ray or beam-halo muons.

The luminosity was measured using the Bethe-Heitler reaction $ep
\rightarrow e \gamma p$ by the luminosity detector which consists of two independent systems. 
In the first system the photons are detected by a lead--scintillator calorimeter placed in the HERA tunnel 107~m 
from the interaction point in the positron-beam direction. The system used in previous ZEUS 
publications~\cite{desy-92-066,*zfp:c63:391,*acpp:b32:2025} was modified by the addition of active filters 
in order to suppress the increased synchrotron radiation background of the upgraded HERA collider.
The second system is a magnetic spectrometer arrangement~\cite{physics-0512153-tmp-43d0e1b7}. A small fraction ($\sim 9$\%) 
of the small-angle energetic photons from the Bethe-Heitler process convert in the 
exit window of the vacuum chamber. 
Electron-positron pairs from the converted photons were bent vertically by a
dipole magnet and detected in tungsten-scintillator calorimeters
located above and below the photon beam at $Z=-104$~m. 
The advantage of the spectrometer system is that it does not suffer from pile-up (multiple interactions at high luminosity)
 and is not sensitive to direct synchrotron radiation, whereas the calorimeter system has higher acceptance.
The fractional uncertainty on the measured luminosity was 3.5\%.

The lepton beam in HERA becomes naturally transversely polarised through the Sokolov-Ternov effect~\cite{sovpdo:8:1203}.
The characteristic build-up time expected for the HERA accelerator is approximately 40~minutes. 
Spin rotators on either side of the ZEUS detector change the 
transverse polarisation of the beam into longitudinal polarisation. The positron beam polarisation was measured using 
two independent polarimeters, the transverse polarimeter (TPOL)~\cite{nim:a329:79} and the longitudinal polarimeter (LPOL)~\cite{nim:a479:334}. 
Both devices exploit the spin-dependent cross section for Compton scattering of circularly polarised photons off positrons to 
measure the beam polarisation. The transverse polarimeter was upgraded in 2001 to provide a fast measurement for every positron 
bunch, and position-sensitive silicon strip and scintillating-fibre detectors were added to investigate systematic 
effects~\cite{tech:pol2000prc}.
The luminosity and polarisation measurements were made over times that were much shorter than the polarisation build-up time.
\section{Monte Carlo simulation}
\label{s:Simu}

Monte Carlo (MC) simulations were used to determine the efficiency for 
selecting events and the accuracy of kinematic 
reconstruction, to estimate the $ep$ background rate and to extrapolate
the measured cross sections to the full kinematic region.
A sufficient number of events were generated to ensure that uncertainties from MC 
statistics were small compared to other uncertainties.

Neutral and charged current DIS events including radiative effects were simulated 
using the {\sc Djangoh}~\cite{proc:hera:1991:1419,*spi:www:djangoh11} generator.
The polarisation dependence of radiative effects in CC DIS, neglected in {\sc Djangoh}, was checked using 
the {\sc Grace}~\cite{kek-92-19,*ptp:138:18} program and found to be negligible.
The hadronic final state was simulated
using the colour-dipole model of {\sc Ariadne} 4.10~\cite{cpc:71:15}
and, as a systematic check, the {\sc Meps} model of
{\sc Lepto} 6.5~\cite{cpc:101:108}.
Both programs use the Lund string model of {\sc Jetset} 7.4~\cite{cpc:39:347,*cpc:43:367,*cpc:82:74}
for the hadronisation.
The photoproduction background was estimated using events 
simulated with {\sc Herwig}~5.9~\cite{cpc:67:465}.
Diffractive NC events were generated using the {\sc Rapgap}~2.08/06~\cite{cpc:86:147} program 
and mixed with the non-diffractive
MC events in order to simulate  the hadronic final state accurately.
Background to the CC signal from 
$W$ production was estimated
using the {\sc Epvec}~1.0~\cite{np:b375:3} generator and background from
the production of charged-lepton pairs was generated using the {\sc Grape}~1.1~\cite{cpc:136:126} program.

The vertex distribution in data is a crucial input to the MC simulation for the
correct evaluation of the event-selection efficiency. Therefore, the $Z$-vertex
distribution used in the MC simulation was determined from a sample of NC DIS
events in which the event-selection efficiency was independent of $Z$.
\section{Kinematic Reconstruction}
\label{s:Rec}

Charged current events are characterised by a large missing transverse momentum, $\PTM$, the magnitude of 
which is calculated as

\begin{equation}
\PTM^2  =  P_x^2 + P_y^2 = 
  \left( \sum\limits_{i} E_i \sin \theta_i \cos \phi_i \right)^2
+ \left( \sum\limits_{i} E_i \sin \theta_i \sin \phi_i \right)^2, \nonumber
  \label{eq:pt}
\end{equation}

where the sum runs over all calorimeter energy deposits $E_i$, (corrected~\cite{epj:c11:427} 
for energy loss in inactive material and other effects in the offline analysis) and
$\theta_i$ and $\phi_i$ are the polar and azimuthal angles
of the calorimeter energy deposit as viewed from the interaction vertex.
The hadronic polar angle, $\gamma_h$, is defined by
$\cos\gamma_h = (\PTM^2 - \delta^2)/(\PTM^2 + \delta^2)$,
where
$\delta = \sum ( E_i - E_i \cos \theta_{i} ) 
= \sum (E-P_z)_{i}$.
In the naive Quark Parton Model,
$\gamma_h$ gives the scattering angle of the struck quark in the lab frame.
The total transverse energy,
$E_T$, is given by
$E_T    = \sum E_i \sin \theta_i$. 
The kinematic variables $x_{\rm JB}$, $y_{\rm JB}$ and $Q^{2}_{\rm JB}$ for charged current events were reconstructed from 
the measured $\PTM$ and $\delta$ using the Jacquet-Blondel method \cite{proc:epfacility:1979:391}.

Neutral current events are characterised by the presence of a high-energy isolated scattered positron in the detector.
It follows from longitudinal momentum conservation that for well measured NC events, $\delta$ peaks at twice the positron beam 
energy or $55 \gev$. 
The hadronic transverse momentum, $P_{T,h}$, and $\delta_h$ were calculated in the same way as the corresponding quantities in CC events,
but excluding energy deposits associated with the scattered positron. The hadronic polar angle, $\gamma_h$, was calculated from $P_{T,h}$, 
and $\delta_h$ in the same way as the CC case. The scattered positron energy, $E^{\prime}_{e}$, and polar angle, $\theta_e$, 
were determined from the energy deposit and matched track of the scattered positron candidate, respectively.
 
The double-angle method~\cite{proc:hera:1991:23,*proc:hera:1991:43} was used to estimate the kinematic variables 
$x_{\rm DA}$, $y_{\rm DA}$ and $Q^{2}_{\rm DA}$ for the neutral current events using the measured values of $\theta_e$ and $\gamma_h$.

\section{Event selection}
\label{s:EvSel}

ZEUS operates a three-level trigger system~\cite{zeus:1993:bluebook,uproc:chep:1992:222}. 
Charged current events were selected using criteria based on missing transverse 
momentum measured by the CAL~\cite{epj:c32:1}.
Neutral current DIS events were selected using criteria based on an energy deposit in the 
CAL consistent with an isolated positron~\cite{pr:d70:052001-tmp-43d0e0ea}. 
 
\subsection{Charged current}

The following criteria were imposed to select CC events and to reject background:

\begin{itemize}
  \item {missing transverse momentum: $\PTM>12 \gev$ was required and, in addition, the missing 
    transverse momentum, excluding the calorimeter cells adjacent to the forward beam hole, $\PTM'$, 
    was required to satisfy  $\PTM'> 10 \gev$. The $\PTM '$ cut strongly suppresses beam-gas events while
    maintaining high efficiency for CC events;}
  \item {primary vertex: events were required to satisfy $| Z_{\rm VTX} | < 50$~cm.
    The $Z$ coordinate of the vertex, reconstructed 
    using the tracking detectors, 
    was required to be consistent with that of an $ep$ interaction. For events 
    with an hadronic angle, $\gamma_h$, of less than $23^\circ$, the vertex $Z$ position was
    reconstructed from the measured arrival time of energy deposits in FCAL~\cite{pl:b316:412}, 
    and the $\PTM$ and $\PTM'$ thresholds were increased to 14 and 12 \gev, respectively;}
  \item {rejection of photoproduction: $\PTM/E_{T}>0.4$ was required for events with $20<\PTM<30 \gev$; 
    $\PTM/E_{T}>0.55$ was
    required for events with $\PTM<20 \gev$. These requirements select events with an azimuthally collimated
    energy flow. In addition, it was required that the angle between the transverse 
    momentum measured using the tracks 
    and that measured by the calorimeter was less than one radian for events with $\PTM<30 \gev$;}
  \item {rejection of NC DIS: NC DIS events in which the scattered positron or the hadronic system 
    is poorly measured can have significant 
    missing transverse momentum. Events with $\delta > 30 \gev$ and an isolated electromagnetic 
    cluster with energy of at least $4\gev$ measured in the calorimeter were rejected;}
  \item {rejection of non-$ep$ background: interactions between one of the beams and the residual 
    gas in the beam pipe or upstream accelerator components can lead to events
    with significant missing transverse momentum. However, for such interactions, 
    the arrival times of energy deposits in the 
    calorimeter are inconsistent with the bunch-crossing time and were used to reject 
    such events. Muon-finding algorithms based on tracking, calorimeter and muon-chamber 
    information were used to reject events caused by cosmic rays or muons in the beam halo.
    In addition, the shape of hadronic showers in the calorimeter was used to reject halo-muon
    events depositing energy in the forward calorimeter. Further details can be found elsewhere~\cite{thesis:kataoka:2005-tmp-41b73725-tmp-43cd33d1-tmp-43cd37be-tmp-43d0e4cd-tmp-44090bcc,thesis:gabareen:2006-tmp-4408d3b9};}
  \item {kinematic region: 
    events were required to satisfy $Q^{2}_{\rm JB}>200 \gev^2$ and $y_{\rm JB}<0.9$. These requirements 
    restricted the event sample to a region where the resolution of the kinematic quantities was
    good and the background was small~\cite{epj:c32:1}.}
\end{itemize}

All events were visually inspected; 12 cosmic-ray and halo-muon events were removed from the negative-polarisation
sample and 8 from the positive-polarisation sample. A total of 158 data events satisfied all criteria in the 
negative-polarisation sample and 311 in the positive-polarisation sample. 

The main background remaining after the selection was photoproduction events, 
the cross section for which is independent of the longitudinal
polarisation of the positron beam. The contamination was estimated from MC to be typically less 
than 1\% but was as high as 5\% in the lowest-$Q^{2}$ bin of the negative-polarisation sample.

\begin{figure}
  \begin{center}
    \includegraphics*[width=.8\textwidth]{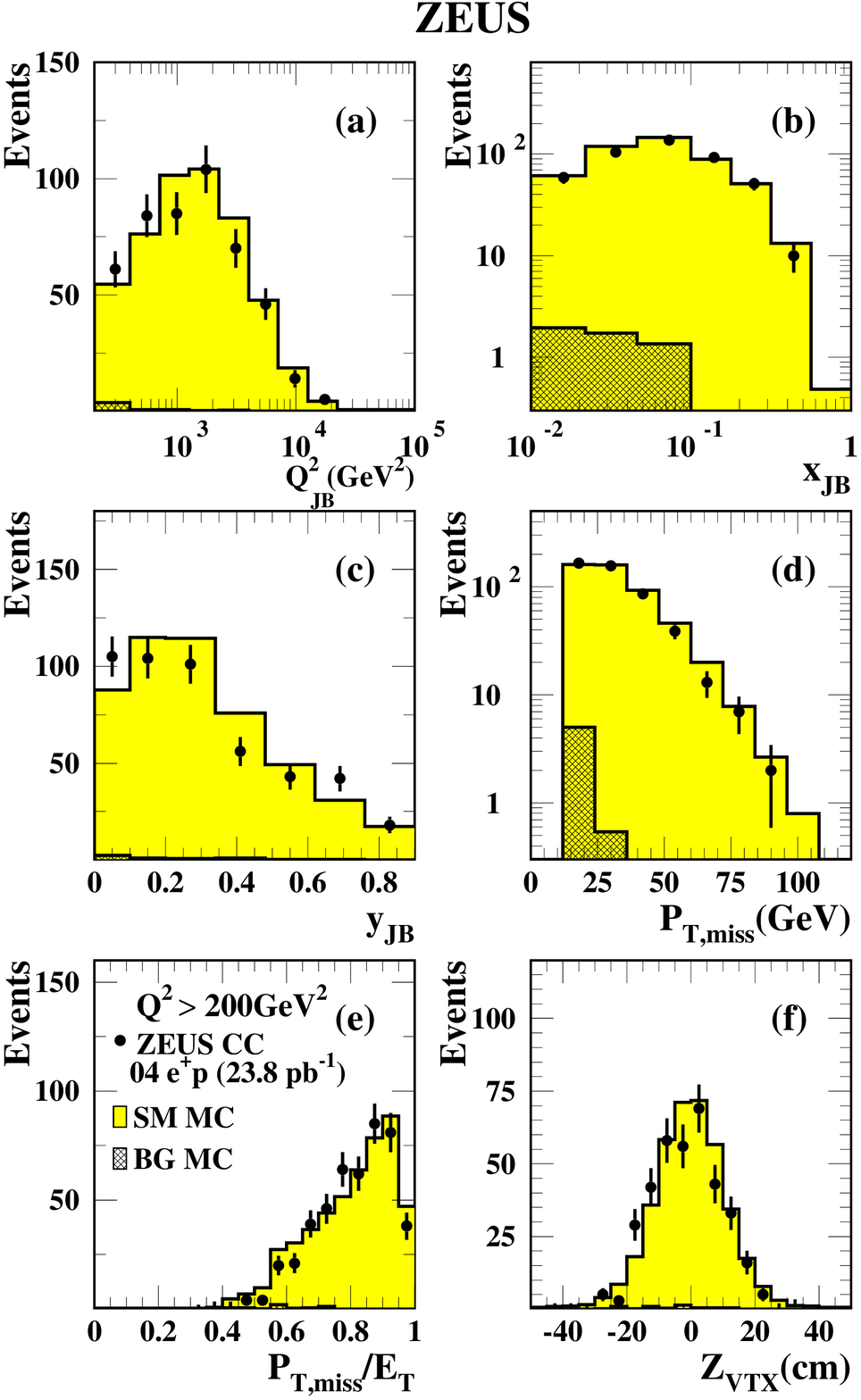}
 \end{center}
  \caption{
    Comparison of the final $e^+ p$ CC data sample (solid points) with the sum of
    the signal and background MC simulations (light shaded histograms). The simulated background events
    are shown as the dark shaded histograms.
    The distributions of (a) $Q^{2}_{\rm JB}$, (b) $x_{\rm JB}$, (c) $y_{\rm JB}$, (d) 
     $\PTM$, (e) $\PTM/E_{T}$ and (f) $Z_{\rm VTX}$, are shown.
    }
  \label{f:cc_ctrl}
\end{figure}

Figure~\ref{f:cc_ctrl} shows a comparison of data and MC distributions for the CC sample. The MC sample, which was weighted to
the measured polarisations and luminosities of the data samples, gives a satisfactory description of the data.

\subsection{Neutral current}

The following criteria were imposed to select NC events:

\begin{itemize}
\item {positron identification: an algorithm which combined information from the energy deposits 
  in the calorimeter with tracks 
   was used to identify scattered positrons. A fiducial-volume cut was applied to
  guarantee that the experimental acceptance was well understood~\cite{pr:d70:052001-tmp-43d0e0ea}. To ensure high purity and reject 
  background, the identified positron was required to have an energy of at least $10 \gev$ and be 
  isolated such that the energy in an $\eta-\phi$ cone of radius 0.8 centred on the positron,
  but not associated with it, was less than $5 \gev$. For events in which a positron was found 
  within the acceptance of the tracking detectors, a track matched to the energy deposit
  in the calorimeter was required. For events with a positron at a smaller polar angle than the 
  acceptance of the tracking detectors, the track requirement was replaced with the requirement that the transverse
  momentum of the positron exceed $30 \gev$;}
\item {primary vertex: events were required to satisfy $| Z_{\rm VTX} | < 50$~cm.
  The $Z$ coordinate of the $ep$ interaction vertex was reconstructed using tracks;}
\item {background rejection: the requirement $38 < \delta < 65 \gev$ was imposed to remove 
  photoproduction and beam-gas
  events, and to reduce the number of events with significant QED initial-state radiation.
  The lower threshold was increased to $44 \gev$ for events which did not have a track matched
  to the positron candidate. To further reduce background from photoproduction, $y$ 
  calculated using the electron method was required to satisfy $y_{e}<0.95$. The net transverse 
  momentum is expected to be small, so, in order to remove cosmic-ray events and beam related
  background events, the quantity $\PTM / \sqrt{E_{T}}$ was required to be less than $4\sqrt{\gev}$, 
  and the quantity $\PTM /E_{T}$ was required to be less than 0.7;}
\item {QED Compton rejection: to reduce the size of the QED radiative corrections, elastic Compton-scattering
  events were rejected. The contribution from deeply-virtual Compton scattering was negligible;}
\item {kinematic region: to avoid regions of phase space in which the MC generator was not valid, the 
  quantity $y_{\rm JB}(1-x_{\rm DA})$ was required to be greater than 0.004. The final event 
  sample was defined by requiring $Q^{2}_{\rm DA}>200 \gev^2$.}
\end{itemize}

\begin{figure}
  \begin{center}
    \includegraphics[width=.8\textwidth]{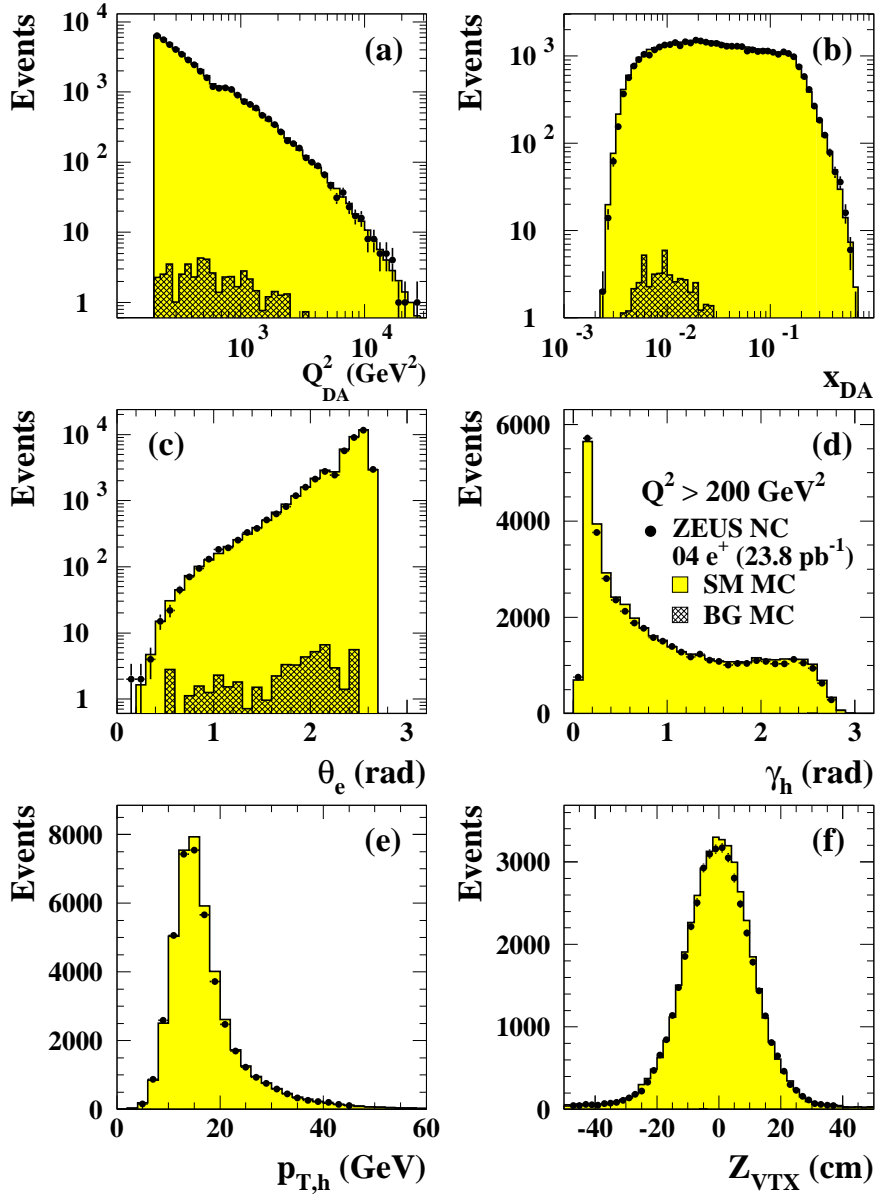}
  \end{center}
  \vskip -5mm
  \caption{
    Comparison of the final $e^+ p$ NC data sample (solid points) with the sum of
    the signal and background MC simulations (light shaded histograms). The simulated background events
    are shown as the dark shaded histograms.
    The distributions of (a) $Q^{2}_{\rm DA}$, (b) $x_{\rm DA}$, (c)  
    $\theta_{e}$, (d)  $\gamma_{h}$, (e)  $P_{T,h}$, and (f) $Z_{\rm VTX}$, are shown.
    }
  \label{f:nc_ctrl}
\end{figure}

A total of 20642 events passed the selection criteria in the negative polarisation sample and 
22395 in the positive polarisation sample. The background contamination, dominated by misidentified photoproduction, was 
typically less than 1\%. Figure~\ref{f:nc_ctrl} shows a 
comparison of data and MC distributions for the NC sample. The MC sample gives a generally 
good description of the data. The effect of the positron fiducial-volume cuts can be seen in the positron angle ($\sim 2.4$~rad)
and $Q^{2}$ ($\sim 600 \gev^{2}$) distributions.

\section{Cross section determination}
 The measured cross section in a particular kinematic bin, for example in $d\sigma/dQ^{2}$, was determined from

 \begin{equation}
 \frac{d\sigma_{\rm Born}}{dQ^{2}}=\frac{N_{\rm data}-N_{\rm bg}}{N_{\rm MC}}\cdot\frac{d\sigma_{\rm Born}^{\rm SM}}{dQ^{2}}, \nonumber
 \end{equation}

 where $N_{\rm data}$ is the number of data events, $N_{\rm bg}$ is the number of background events estimated from the MC simulation and $N_{\rm MC}$ is the number of signal MC events. The SM prediction $d\sigma_{\rm Born}^{\rm SM}/dQ^{2}$ was evaluated in the on-shell scheme using the PDG~\cite{pl:b592:1} values for the electroweak parameters and the CTEQ5D PDFs~\cite{epj:c12:375}. Consequently, the acceptance, as well as the bin-centring and 
radiative corrections were all taken from the MC simulation. The radiative corrections define the measured cross section to have only tree-level QED and electroweak contributions.
A similar procedure was used for $d\sigma/dx$ and $d\sigma/dy$.

The major sources of systematic uncertainty
in the CC cross sections come from the uncertainties in calorimeter energy scale and
 the parton-shower scheme. 
The former was estimated using a method detailed in previous publications~\cite{pl:b539:197,epj:c32:1} for the NC data sample. 
The resulting shifts in the cross sections 
were typically less than
$10\%$, but increased to $20\%$ in the highest $Q^2$ bin and
$30\%$ in the highest $x$ bin. 

To estimate the sensitivity of the results to the details of the simulation of
the hadronic final state,
the {\sc Lepto} {\sc Meps} model was used instead of the 
{\sc Ariadne} model for calculating the acceptance corrections.
The largest effects of $\sim 5\%$ were observed in the highest $Q^2$ and $x$ bins. 

The uncertainty in the small contribution from photoproduction was estimated by fitting a 
linear combination of the $\PTM /E_{T}$ distributions of the 
signal and the background MC samples to the corresponding distribution in the data, 
allowing the normalisation of the photoproduction MC events to vary. No cut 
on $\PTM /E_{T}$ was applied for this check. Varying the normalisation of the photoproduction events by 
the uncertainty in the fit of $\pm 30\%$ resulted in changes of the measured cross sections within $\pm 3\%$.

The systematic uncertainties of the selection cuts were estimated by varying
the threshold value of each selection cut independently   
by around 10\%, which is a reasonable match to the resolution.
The resulting shifts in the cross sections were typically within $\pm 5\%$.

A major source of systematic uncertainty
in the NC cross section came from the uncertainty in the parton-shower scheme, which gave changes in the 
cross section of typically within $\pm 2\%$ but up to 4\% at high $Q^{2}$. Uncertainty in the electromagnetic 
energy scale was estimated by varying the energy scale by $\pm 1\%$. However, due to the use of the 
double-angle reconstruction, the resulting shifts in the cross section were typically $<0.5\%$.
The systematic effects of the selection cuts were estimated by varying
the threshold value of each selection cut independently   
by values commensurate with the resolutions.
The resulting shifts in the cross sections were typically within $\pm 1\%$.

The individual uncertainties were added in quadrature separately
for the positive and negative deviations from the
nominal cross-section values to obtain the total systematic uncertainty. 
The uncertainty in the measured polarisation, $\delta P_{e}/P_{e}$, was 1.6\% using the LPOL 
and 3.5\% using the TPOL. The choice of polarimeter measurement was made to maximise 
the available luminosity for the analysis, while minimising the uncertainty in the measured polarisation, on a run-by-run basis.

The relative uncertainty in the measured luminosity
of 3.5\% was not included in the total uncertainty shown in the differential cross-section figures.

\section{Results}
\label{ss:Results}

In the following, measurements of total cross sections and differential cross sections in
$x$, $y$ and $Q^{2}$ for the charged current reaction are presented. In addition, cross
sections differential in $Q^{2}$ were measured for the neutral current reaction.

\begin{figure}
  \begin{center}
    \includegraphics[width=.8\textwidth]{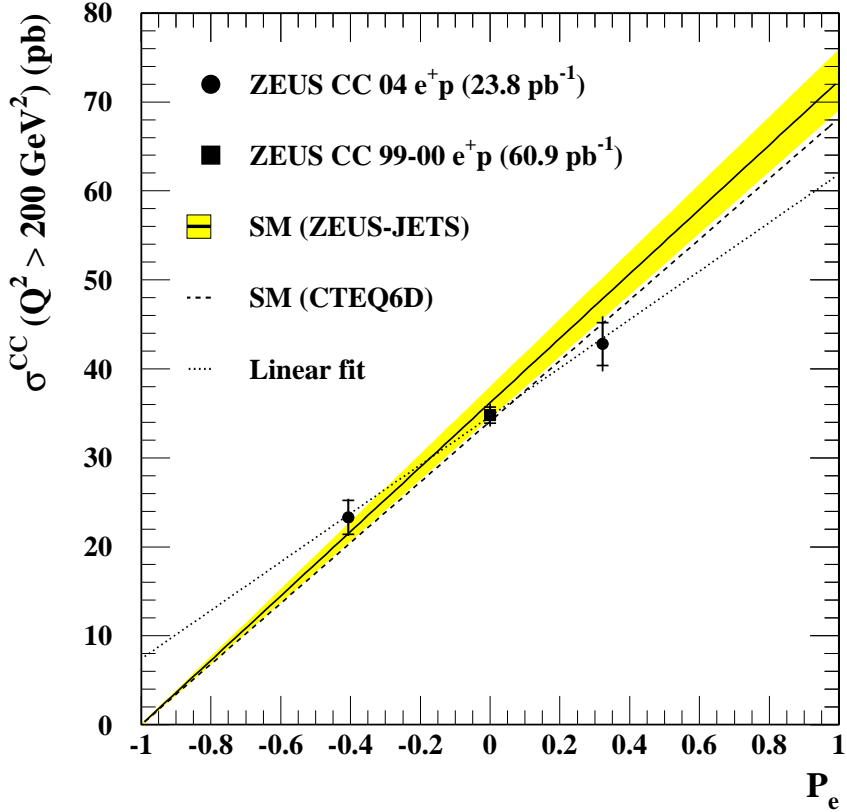}
  \end{center}
  \caption{ 
    The total cross section for $e^+ p$ CC DIS with $Q^{2}>200 \gev ^{2}$ as a function of the longitudinal 
    polarisation of the positron beam. The full (dashed) line shows the prediction of the SM evaluated using
    the ZEUS-JETS (CTEQ6D) PDFs and the shaded band indicates the uncertainty on the cross section from the 
    ZEUS-JETS fit. The dotted line shows the result of the linear fit to the data described in the text.
    Horizontal error bars representing the uncertainty on the measured polarisation are included but are too small
    to be visible.
    }
  \label{f:cctot}
\end{figure}

The total cross sections for $e^+ p$ CC DIS in the kinematic region $Q^{2}>200 \gev^{2}$ are

\begin{equation}
\sigma^{\rm CC}(P_{e}=0.32 \pm 0.01)=42.8 \pm 2.4 ({\rm stat.}) \pm 1.9 ({\rm syst.})~{\rm 
pb}, \nonumber
\end{equation}

and 

\begin{equation}
\sigma^{\rm CC}(P_{e}=-0.41 \pm 0.01)=23.3 \pm 1.9 ({\rm stat.}) \pm 1.0 ({\rm syst.})~{\rm
 pb}. \nonumber 
\end{equation}

including the uncertainty from the measured luminosity. The total cross section is shown as a function of 
the longitudinal polarisation of the positron beam in Fig.~\ref{f:cctot}, including the unpolarised ZEUS 
measurement from the 1999-2000 data~\cite{epj:c32:1}. 
The data are compared to the Standard Model prediction evaluated using the ZEUS-JETS~\cite{epj:c42:1-tmp-4396fb15}
and CTEQ6D~\cite{jhep:07:012} PDFs. 
The SM prediction describes the data well. A linear fit to the data yields an extrapolated value of

\begin{equation}
\sigma^{\rm CC}(P_{e}=-1)=7.4 \pm 3.9 ({\rm stat.}) \pm 1.2 ({\rm syst.})~{\rm pb}, \nonumber 
\end{equation}

with $\chi^2 = 0.1$, consistent within two standard deviations 
with the absence of right-handed charged currents in the SM. In the fit,
the systematic uncertainties of the two polarised data points were considered fully correlated and the uncertainities
in the measured polarisation fully anti-correlated. The systematic uncertainty in the unpolarised data point was considered
to be uncorrelated with the polarised points.

\begin{figure}
  \begin{center}
    \includegraphics[width=.8\textwidth]{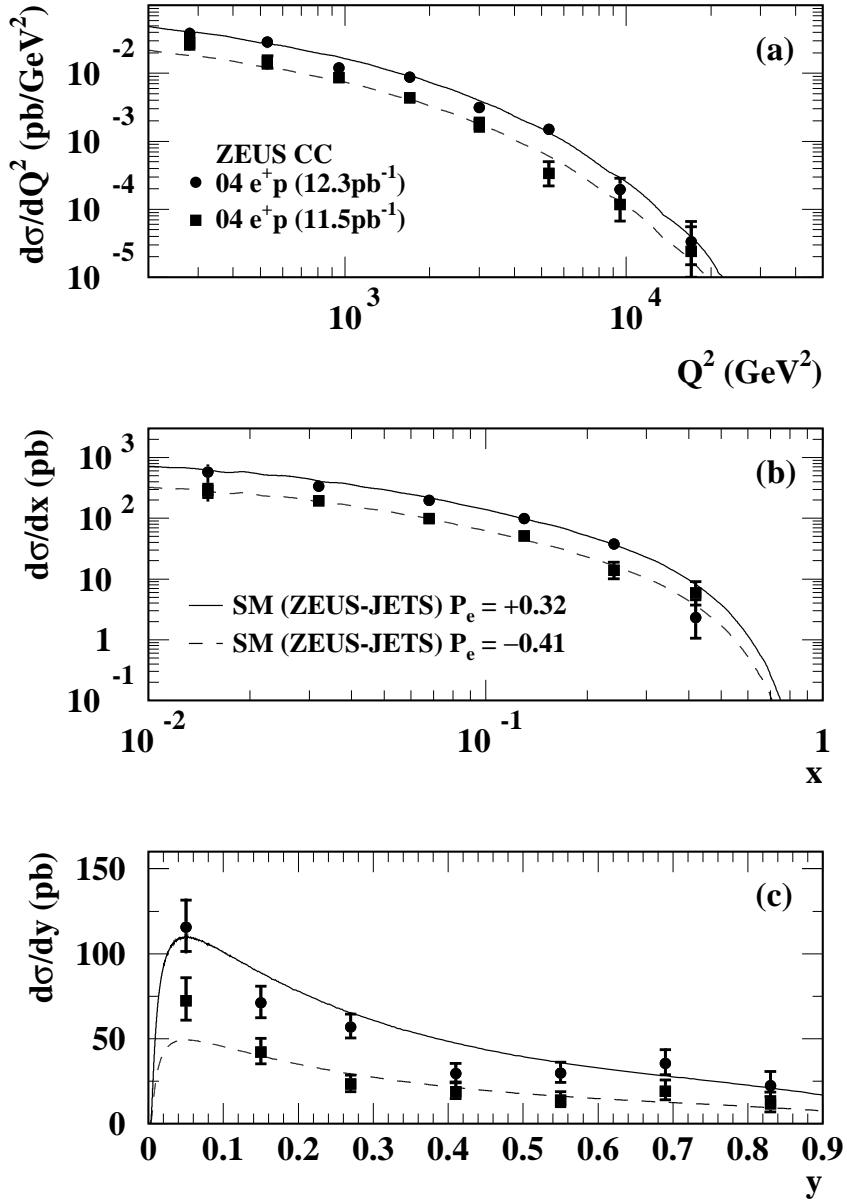}
  \end{center}
  \caption{ 
    The $e^+ p$ CC DIS cross-sections (a) $d\sigma/dQ^{2}$, (b) $d\sigma/dx$ and 
    (c) $d\sigma/dy$. The circles (squares) represent data points for the positive (negative)
    polarisation measurements and the curves show the predictions of the SM evaluated using
    the ZEUS-JETS PDFs.
    }
  \label{f:single_cc}
\end{figure}

The single-differential cross-sections $d\sigma/dQ^{2}$, $d\sigma/dx$ and $d\sigma/dy$ for charged current DIS 
are shown in Fig.~\ref{f:single_cc}. A clear difference
is observed between the measurements for positive and negative longitudinal polarisation, which is independent of the kinematic variables.
The effects are well described by the SM evaluated using the ZEUS-JETS PDFs.

\begin{figure}
  \begin{center}
    \includegraphics[width=.8\textwidth]{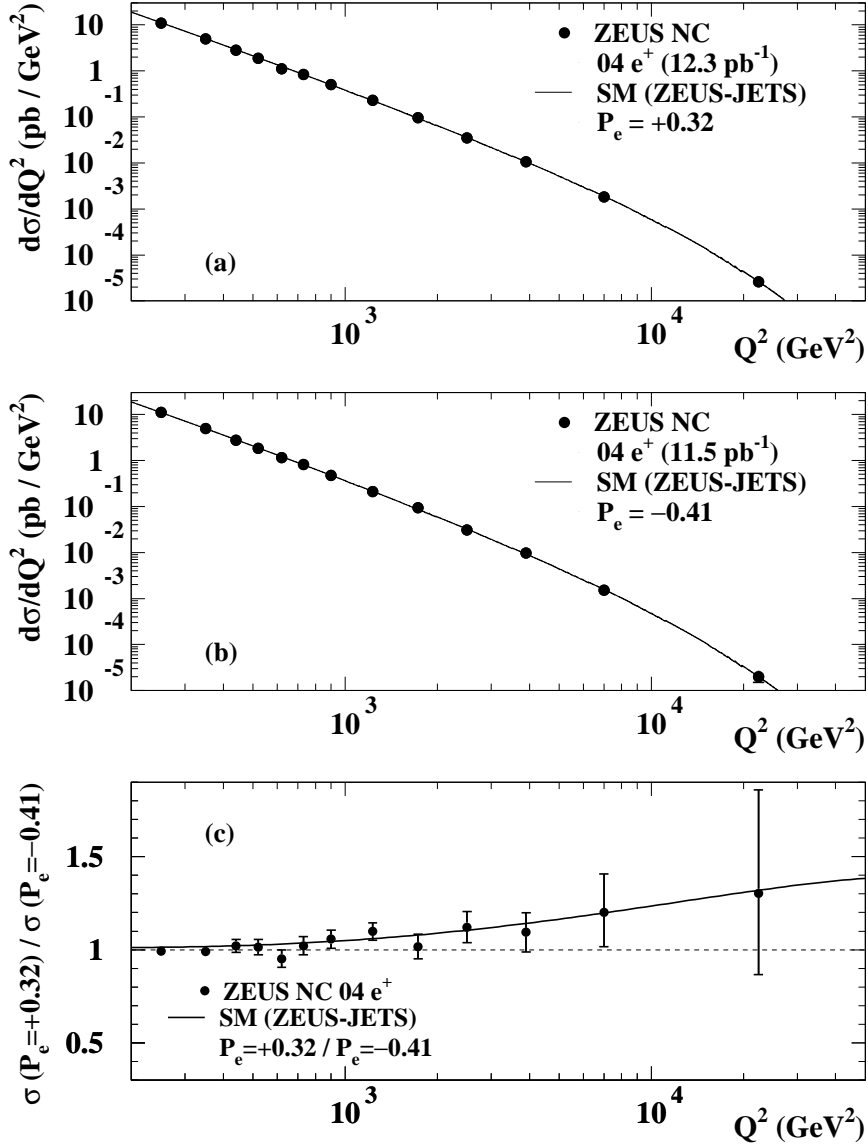}
  \end{center}
  \caption{ 
    The $e^+ p$ NC DIS cross-section $d\sigma/dQ^{2}$ for (a) positive polarisation data,
    (b) negative polarisation data and (c) the ratio of the two. The full lines
    show the predictions of the SM evaluated using the ZEUS-JETS PDFs and the dashed line the prediction
    with no dependence on the longitudinal polarisation of the positron beam.
    }
  \label{f:single_nc}
\end{figure}

Figure~\ref{f:single_nc} shows the differential cross-section $d\sigma/dQ^{2}$ for NC DIS for positive and negative 
longitudinal polarisations and the ratio of the two cross sections. Only
statistical uncertainties were considered when taking the ratio  
of the positively and negatively polarised cross sections. The measurements 
are well described by the SM evaluated using the ZEUS-JETS PDFs and are consistent with
the expectations of the electroweak Standard Model for polarised NC DIS. A $\chi^2$ test for the 
$Q^2 > 1000 \gev^2$ data points yields
$\chi^2 = 0.3$ per data point for the SM and 1.5 for no polarisation dependence.

\section{Summary}
\label{s:summary}

The cross sections for charged and neutral current deep inelastic scattering 
in $e^{+}p$ collisions with a longitudinally polarised positron beam have been measured. 
The measurements are the first from the ZEUS collaboration in the second phase of HERA operation and are 
based on data corresponding to an integrated 
luminosity of $23.8 \pb^{-1}$ collected in 
2004 at a centre-of-mass energy of 318~\gev. The cross sections for 
$e^+p$ charged  current deep inelastic scattering are different for positive 
and negative values of the positron beam longitudinal polarisation. In addition, 
single differential cross sections are presented for charged and neutral 
current deep inelastic scattering in the kinematic region 
$Q^{2}>200 \gev^{2}$. The measured cross sections are well described by the 
predictions of the Standard Model. A fit to the cross-section measurements yields
\mbox{$\sigma^{\rm CC}(P_{e}=-1)=7.4 \pm 3.9 ({\rm stat.}) \pm 1.2 ({\rm syst.})~{\rm pb}$}, which is
within two standard deviations of the prediction of the Standard Model of zero. 

\section*{Acknowledgements}

We are grateful to the DESY directorate for their strong
support and encouragement. We thank the HERA machine
group whose outstanding efforts
resulted in the successful upgrade of the HERA accelerator
which made this work possible.  We also thank the HERA
polarimeter group for providing the measurements of the
lepton-beam polarisation.  The design, construction and
installation of the ZEUS detector has been made possible by
the efforts of many people not listed as authors. It is
a pleasure to thank H. Spiesberger and T. Abe for useful
discussions.

\vfill\eject

{
\def\bibname{\Large\bf References}
\def\refname{\Large\bf References}
\pagestyle{plain}
\ifzeusbst
  \bibliographystyle{./BiBTeX/bst/l4z_default}
\fi
\ifzdrftbst
  \bibliographystyle{./BiBTeX/bst/l4z_draft}
\fi
\ifzbstepj
  \bibliographystyle{./BiBTeX/bst/l4z_epj}
\fi
\ifzbstnp
  \bibliographystyle{./BiBTeX/bst/l4z_np}
\fi
\ifzbstpl
  \bibliographystyle{./BiBTeX/bst/l4z_pl}
\fi
{\raggedright
\bibliography{./BiBTeX/user/syn.bib,%
              ./BiBTeX/bib/l4z_articles.bib,%
              ./BiBTeX/bib/l4z_books.bib,%
              ./BiBTeX/bib/l4z_conferences.bib,%
              ./BiBTeX/bib/l4z_h1.bib,%
              ./BiBTeX/bib/l4z_misc.bib,%
              ./BiBTeX/bib/l4z_old.bib,%
              ./BiBTeX/bib/l4z_preprints.bib,%
              ./BiBTeX/bib/l4z_replaced.bib,%
              ./BiBTeX/bib/l4z_temporary.bib,%
              ./BiBTeX/bib/l4z_zeus.bib}}
}
\vfill\eject

%
%
\newpage

%
%
\end{document}